\documentstyle[12pt]{article}
\begin{document}
\large
\begin{flushright}{UT-831 \\
 September ,1998
}
\end{flushright}
\vfil
\vfil
\Large
\begin{center}
Comment  on anomaly matching in N=1 supersymmetric QCD
\\
\vfil
Kazuo Fujikawa
\\

Department of Physics, University of Tokyo\\
Bunkyo-ku, Tokyo 113, Japan
\vfil
\vfil
Abstract

\end{center}
\normalsize
An attempt is made at a systematic approach to anomaly matching problem 
in non-Abelian electric-magnetic duality in  $N=1$
 supersymmetric QCD. A strategy we employ is somewhat analogous to   anomaly analyses   in grand unified models where the anomaly cancellation becomes 
more 
transparent if one embeds $SU(5)$ multiplets into a multiplet of (anomaly-free) $SO(10)$. 
A  complication arises in the treatment of $U^{AF}_{R}(1)^{3}$ matching where
 $U^{AF}_{R}(1)$ is anomaly-free $R$ symmetry. It is  noted that a  relatively 
systematic analysis of the anomaly matching is possible if one  considers the 
formal breaking sequence of color gauge  symmetry: $SU(N_{f})_{c}\rightarrow SU(N_{c})_{c}\times SU(\tilde{N}_{c})_{c}$ with $N_{f}= N_{c}+ \tilde{N}_{c}$ , where  $N_{f}$ stands for the number of massless quarks.\\
\\
\newpage
\section{Introduction}
The  analyses of non-Abelian electric-magnetic duality of $N=1$ supersymmetric QCD have been initiated  by 
Seiberg [1][2]. As for reviews, see Refs.[3][4]. The basic ingredients 
of the analyses are holomorphicity, decoupling, and the 't Hooft anomaly matching 
condition[5]. 

In this note we comment on some aspects of anomaly matching. 
We concentrate on anomaly matching  , without paying attention
to other important physical inputs such as holomorphicity, superpotential and decoupling, mainly because the anomaly matching provides a very definite mathematical framework: Once one finds a solution to anomaly matching, one has  a good starting point of analysis and one may then exercise one's imagination about the possible physical meaning of the obtained solution.  
Our motivation for this analysis is to find  a more systematic approach to the anomaly matching problem. A strategy we employ is somewhat analogous to the  
anomaly analyses in the conventional grand unification models. For example, 
the anomaly cancellation in the $SU(5)$ scheme is rather miraculous, but if 
one embeds all the multiplets appearing in the $SU(5)$ model into the (anomaly-free) $SO(10)$
model, the anomaly cancellation becomes more systematic and transparent [6].     

However, the present approach as it stands is more involved.  One of the  main reasons is the subtle property of the anomaly free $U_{R}^{AF}(1)$ 
symmetry related to  R-symmetry. From a supersymmetry view point, the $U_{R}^{AF}(1)$ charge  may be regarded as condensed in the vacuum in the sense that the constant SUSY transformation
parameter carries a non-trivial $U_{R}^{AF}(1)$ charge by definition. Stated
differently, the $U_{R}^{AF}(1)$ charge does not commute with the basic generators of $N=1$ supersymmetry though the $U_{R}^{AF}(1)$ charge  commutes
with the Hamiltonian. 
Also, if one decouples one of the massless quarks by giving a mass to  it, 
the entire $U_{R}^{AF}(1)$ charge assignment of the rest of the quarks is 
reshuffled. In this respect, $U_{R}^{AF}(1)$ is quite different from other 
global symmetry such as the flavor symmetry 
$SU(N_{f})_{L}\times SU(N_{f})_{R}$.  

It turns out to be  relatively easy to  achieve an intuitive understanding of all the anomaly matching except for  the triangle $(U_{R}^{AF}(1))^{3}$.  We thus 
suggest to examine the solutions obtained by dropping  the requirement of $(U_{R}^{AF}(1))^{3}$  anomaly matching tentatively and by requiring only the anomaly matching 
{\em linear} in $U_{R}^{AF}(1)$, which is equivalent to imposing the existence of conserved $U_{R}^{AF}(1)$ current (with spurious ``leptons'') without gauging it. The physical 
relevance of gauging  $U_{R}^{AF}(1)$ , which does not commute with the generators of $N=1$ supersymmetry, will be commented on later.

\section{Minimal model with $SU(N)_{c}$ gauge theory}
Following  the analyses in Refs.[1] and [2], we consider the fermion contents of $N=1$ supersymmetric QCD with color $SU(N_{c})_{c}$ for $N_{c}< N_{f}$  by
 denoting  the quantum numbers related to $SU(N_{c})_{c}\times SU(N_{f})_{L}
\times SU(N_{f})_{R} \times U_{B}(1) \times U_{R}^{AF}(1)$ in this order  
\begin{eqnarray}
\Psi_{Q} &:& (N_{c})\ \ (N_{f}, 1, 1, -\frac{N_{c}}{N_{f}})\nonumber\\
\Psi_{\bar{Q}}&:& (\bar{N}_{c})\ \ (1, \bar{N}_{f},- 1, -\frac{N_{c}}{N_{f}})\nonumber\\
\lambda_{N_{c}}&:& (N_{c}^{2}-1)\ \ (1, 1, 0, 1)
\end{eqnarray}
where we write  only the (left-handed) fermion components of quark scalar multiplets,$Q$ and $\bar{Q}$,  and the gaugino $\lambda_{N_{c}}$ of the 
gauge vector multiplet. Here $U(1)_{B}$ stands for the baryon number. We classify $\Psi_{\bar{Q}}$ as $\bar{N}_{f}$ of $SU(N_{f})_{R}$. 
The above multiplet is dual to the (magnetic) $N=1$ supersymmetric theory with $SU(\tilde{N}_{c})_{c}$ gauge 
symmetry given by (in the same notation)
\begin{eqnarray}
\Psi_{q} &:& (\bar{\tilde{N}}_{c})\ \ (\bar{N}_{f}, 1, \frac{N_{c}}{\tilde{N}_{c}}, -\frac{\tilde{N}_{c}}{N_{f}})\nonumber\\
\Psi_{\bar{q}}&:& (\tilde{N}_{c})\ \ (1, N_{f}, -\frac{N_{c}}{\tilde{N}_{c}}, -\frac{\tilde{N}_{c}}{N_{f}})\nonumber\\
\lambda_{\tilde{N}_{c}}&:& (\tilde{N}_{c}^{2}-1)\ \ (1, 1, 0, 1)\nonumber\\
\Psi_{T}&:& (1)\ \ (N_{f}, \bar{N}_{f}, 0, \frac{\tilde{N}_{c}- N_{c}}{N_{f}})
\end{eqnarray}
with $N_{c} + \tilde{N}_{c} = N_{f}$. $\Psi_{T}$ stands for the fermion 
component of a meson scalar multiplet $T$ formed of $Q\bar{Q}$, and $q$ and $\bar{q}$ stand for quarks in the dual theory. We assign $\Psi_{q}$ and $\Psi_{\bar{q}}$ to $(\bar{\tilde{N}}_{c})$ and $ (\tilde{N}_{c})$ of color 
$SU(\tilde{N}_{c})$, respectively, for later convenience by departing from the 
 convention in the original reference [2], but this does not change physics. 

\subsection{$SU(N_{c})_{c}\times SU(N_{f}-N_{c})_{c}\rightarrow SU(N_{f})_{c}$}
Instead of comparing the anomalies associated with the global symmetries 
$SU(N_{f})_{L}
\times SU(N_{f})_{R} \times U_{B}(1) \times U_{R}^{AF}(1)$ in the above mutually dual multiplets directly, we here propose to compare the anomalies of the set of fields (with $N_{f}=N_{c}+\tilde{N}_{c}$ )
\begin{eqnarray}
&&\left. \begin{array}{ccc}
\Psi_{Q} &:& (N_{c}) (N_{f}, 1, 1, -\frac{N_{c}}{N_{f}})\\  
\bar{\Psi}_{q} &:& (\tilde{N}_{c})(N_{f}, 1,- \frac{N_{c}}{\tilde{N}_{c}}, \frac{\tilde{N}_{c}}{N_{f}})
\end{array}\right\}\rightarrow (N_{f})(N_{f}, 1, 0, \frac{\tilde{N}_{c}-N_{c}}{N_{f}})\nonumber\\
&&\left. \begin{array}{ccc}
\Psi_{\bar{Q}} &:& (\bar{N}_{c}) (1, \bar{N}_{f}, -1, -\frac{N_{c}}{N_{f}})\\  
\bar{\Psi}_{\bar{q}} &:& (\bar{\tilde{N}}_{c})(1, \bar{N}_{f}, \frac{N_{c}}{\tilde{N}_{c}}, \frac{\tilde{N}_{c}}{N_{f}})
\end{array}\right\}\rightarrow (\bar{N}_{f})(1, \bar{N}_{f}, 0, \frac{\tilde{N}_{c}-N_{c}}{N_{f}})\nonumber\\
&&\left. \begin{array}{ccc}
\lambda_{N_{c}} &:& (N_{c}^{2}-1) (1, 1, 0, 1)\\  
\bar{\lambda}_{\tilde{N}_{c}} &:& (\tilde{N}_{c}^{2}-1)(1, 1, 0, -1)
\end{array}\right\}\rightarrow (N_{f}, \bar{N}_{f})(1, 1, 0,- \frac{\tilde{N}_{c}-N_{c}}{N_{f}})\nonumber\\
\end{eqnarray}
with those of 
\begin{equation}
\Psi_{T}: (1) (N_{f}, \bar{N}_{f}, 0, \frac{\tilde{N}_{c}-N_{c}}{N_{f}})\rightarrow (1) (N_{f}, \bar{N}_{f}, 0, \frac{\tilde{N}_{c}-N_{c}}{N_{f}})
\end{equation}

Namely, we move the set of fields,  $\Psi_{q}, \Psi_{\bar{q}}$ and $ \lambda_{\tilde{N}_{c}}$, from one side of the duality relation to the other; at the same time, we replace all the moved fields by their ``anti-fields'', which are
defined by reversing all the quantum numbers {\em including} $U^{AF}_{R}(1)$
charge. Note that this ``anti-field'' differs from the physical charge conjugated field since chirality is {\em not} reversed in this definition. ( We here 
utilize the freedom of an overall constant factor of $R$ - charge assignment in a {\em given } theory.)
The fact that this re-arrangement does not change the anomaly matching 
condition is understood in the path integral approach[7], for example. We then multiply the path integral 
\begin{equation}
\int d\bar{\mu}\exp [ i\int \bar{{\cal L}} (\Psi_{q}, \Psi_{\bar{q}}, 
\lambda_{\tilde{N}_{c}})d^{4}x ]
\end{equation}
to both sides of the duality relation, where $\bar{{\cal L}}$ stands for the 
Lagrangian for ``anti-fields'' defined above. Since the path integral $\int d\mu d\bar{\mu}\exp
[i\int (\bar{{\cal L}} + {\cal L}) d^{4} x ]$ is anomaly- free for all the global symmetries, we maintain the equivalence of the anomaly matching condition 
by this procedure. From an anomaly matching view point , it is simpler to move all the Lagrangians
to the one-side of the duality ralation. In this case, the anomaly matching becomes equivalent to  anomaly cancellation without introducing spurious ``leptons''. We thus have a
better analogy to the anomaly cancellation in grand unification schemes. In 
fact, the global symmetry $SU(N_{f})_{L}\times SU(N_{f})_{R}\times U_{B}(1)$
behaves  analogously. However, the symmetry  $U^{AF}_{R}(1)$ behaves 
 differently  in many respects such as a drastic 
reshuffling of quantum number assignment.
The reason why we compare (3) and (4) will become clear later.

In the above relations in eq.(3), we first classify the fields according to 
their representation of $SU(N_{f})_{L}\times SU(N_{f})_{R}$, which turns out to be a sole solid classification symmetry in the present model,  and then combine the fields pair-wise, for example, $\Psi_{Q}+ \bar{\Psi}_{q}$ into a multiplet of the color gauge symmetry of $SU(N_{f})_{c}$. Note that $N_{f} = N_{c} + \tilde{N}_{c}$. 
From a superfield view point, this operation symbolically means
\begin{eqnarray}
&&\int {\cal L}(\Psi_{Q}, \Psi_{\bar{Q}}, \lambda_{N_{c}})(x,\theta_{1},\bar{\theta_{1}})d^{4}xd^{4}\theta_{1} + \int {\cal L}(\bar{\Psi}_{q}, \bar{\Psi}_{\bar{q}}, \bar{\lambda}_{\tilde{N}_{c}})(x,\theta_{2},\bar{\theta_{2}})d^{4}xd^{4}\theta_{2}\nonumber\\
&& \rightarrow
\int {\cal L}(\Psi_{Q}+ \bar{\Psi}_{q}, \Psi_{\bar{Q}}+ \bar{\Psi}_{\bar{q}} , \lambda_{N_{c}+\tilde{N}_{c}})(x,\theta,\bar{\theta})d^{4}xd^{4}\theta
\end{eqnarray}
The Grassmann parameters $\theta_{1}$ and $\theta_{2}$ are transformed {\em
oppositely} under $R$-symmetry, as is indicated in the left-hand side of (3),  by using the freedom of an overall constant factor for R-charge assignment within a given theory. This operation (6)
causes a complete reshuffling of $U^{AF}_{R}(1)$ charge assignment in the 
combined theory in (3).   
 
In (3) the number of quark freedom is kept fixed, but the gauge
degrees of freedom together with gaugino freedom are changed. Within this pair-wise combination, we can match anomalies associated with 
$U_{B}(1)(SU(N_{f})_{L})^{2}$,  $U_{B}(1)(SU(N_{f})_{R})^{2}$,  $(SU(N_{f})_{L})^{3}$,  $(SU(N_{f})_{R})^{3}$,  $ U^{AF}_{R}(1)(SU(N_{f})_{L})^{2}$,\\  $U^{AF}_{R}(1)(SU(N_{f})_{R})^{2}$,  $(U_{B}(1))^{2}U^{AF}_{R}(1)$ and  $U^{AF}_{R}(1)$ - gravitational anomaly (i.e., the freedom counting) except for $U^{AF}_{R}(1)^{3}$.  
For example, for the combination of $\Psi_{Q} + \bar{\Psi}_{q}$ we have the coefficients of anomalies:
\begin{eqnarray}
&&U_{B}(1)(SU(N_{f}))^{2} : N_{c}\times 1 + \tilde{N}_{c}\times (-\frac{N_{c}}{\tilde{N}_{c}}) =0 \rightarrow 0\nonumber\\
&&(SU(N_{f})_{L})^{3} : N_{c} + \tilde{N}_{c} = N_{f} \rightarrow N_{f}\nonumber\\
&&U^{AF}_{R}(1)(SU(N_{f}))^{2} : N_{c}(-\frac{N_{c}}{N_{f}}) + \tilde{N}_{c}(\frac{\tilde{N}_{c}}{N_{f}}) \rightarrow N_{f}(\frac{\tilde{N}_{c}- N_{c}}{N_{f}})\nonumber\\
&&U^{AF}_{R}(1)(U_{B}(1))^{2} : N_{c}(-\frac{N_{c}}{N_{f}}) \times 1 +
\tilde{N}_{c}\frac{N_{c}}{N_{f}}(-\frac{N_{c}}{\tilde{N}_{c}})^{2}= 0 \rightarrow 0\nonumber\\
&& U^{AF}_{R}(1)-gravitational : N_{c}(-\frac{N_{c}}{N_{f}}) + \tilde{N}_{c}
(\frac{\tilde{N}_{c}}{N_{f}}) \rightarrow N_{f}(\frac{\tilde{N}_{c}- N_{c}}
{N_{f}})
\end{eqnarray}
Similarly one can confirm the anomaly matching for other combinations in eq.(3).
 
The anomaly of $U^{AF}_{R}(1)$ related to the instanton of color gauge fields
becomes
\begin{eqnarray}
\Psi_{Q} + \bar{\Psi}_{q}&:& \frac{N_{f}}{2}(-\frac{N_{c}}{N_{f}}){\cal P}(SU(N_{c})) +
\frac{N_{f}}{2}(\frac{\tilde{N}_{c}}{N_{f}}){\cal P}(SU(\tilde{N}_{c}))\nonumber\\
&&\rightarrow \frac{N_{f}}{2}(\frac{\tilde{N}_{c}- N_{c}}{N_{f}}){\cal P}(SU(N_{f})
\end{eqnarray}
where ${\cal P}(SU(N_{c}))$, for example,  stands for the Pontryagin index for $SU(N_{c})$ color 
gauge fields. A similar relation for $\Psi_{\bar{Q}} + \bar{\Psi}_{\bar{q}}$
together with the one for 
\begin{equation}
\lambda_{N_{c}} + \bar{\lambda}_{\tilde{N}_{c}} : N_{c}{\cal P}(SU(N_{c})) - \tilde{N}_{c}{\cal P}(SU(\tilde{N}_{c})) \rightarrow N_{f}(\frac{N_{c}-\tilde{N}_{c}}{N_{f}}){\cal P}(SU(N_{f}))
\end{equation}
shows that the assignment of $U^{AF}_{R}(1)$ charge  is consistent, namely, the sum of twice of eq.(8) and eq.(9) vanishes in the both sides of correspondence. The color singlet component in the right-hand side of $ \lambda_{N_{c}} + \bar{\lambda}_{\tilde{N}_{c}}$ in eq.(3) does not contribute in this calculation.
It is confirmed that $U^{AF}_{R}(1)$ charge assignment has a {\em unique} solution in (3) if one assumes the appearance of a 
minimum set of fields as in (3), namely,  if we allow no degenerate fields with
respect to   $SU(N)_{c}\times SU(N_{f})_{L}
\times SU(N_{f})_{R} \times U_{B}(1) $
which can be distinguished only by different  $U^{AF}_{R}(1)$ charges.    
The anomaly $( U^{AF}_{R}(1))^{3}$ is not matched within the pair-wise combination , although overall it is matched in eq.(3), and thus it is non-trivial from the present view point. 
 
If one looks at the right-hand side of the correspondence in eqs.(3) and (4) and compare the multiplets appearing there, 
all the anomaly matching including $( U^{AF}_{R}(1))^{3}$ is manifest, if one 
remembers $N_{f} = N_{c} + \tilde{N}_{c}$. The anomaly-free condition of $U^{AF}_{R}(1)$ related to instantons is also manifestly
satisfied. As for calculational rules of anomalies, see (7) $\sim$ (9). Note that the assignment of $U^{AF}_{R}(1)$ charge to all the fields is arbitrary up to a common overall constant. [ The absolute normalization of $R$ charge becomes relevant  when one analyzes the superconformal theory
and the relation such as $d\geq \frac{3}{2}|R|$, since one has to assign a 
proper (length) dimension to the Grassmann parameter in such an analysis.]
By this  embedding into  a larger gauge group, one can understand the anomaly matching condition in the original duality relation between (1) and (2) in a more systematic way. 
 
\subsection{$SU(N_{f})_{c}\rightarrow SU(N_{c})_{c}\times SU(N_{f}-N_{c})_{c}$}
We now look at the above correspondence in eqs.(3) and (4) in a reversed order, namely  from the right-hand side to the left-hand side. In this case, the extra color singlet component of the 
gluino field of $SU(N_{f})_{c}$ in (3) is identified as an ``anti-field'' of 
baryon in the duality relation for the case of $N_{c} = N_{f}, N=1$ 
supersymmetric  QCD[1]. Namely, in the notation of $SU(N)_{c}\times SU(N_{f})_{L}
\times SU(N_{f})_{R} \times U_{B}(1) \times U_{R}^{AF}(1)$ 
\begin{eqnarray}
\Psi_{Q} &:& (N_{f})\ \ (N_{f}, 1, 0, \frac{\tilde{N}_{c}- N_{c}}{N_{f}})\nonumber\\
\Psi_{\bar{Q}}&:& (\bar{N}_{f})\ \ (1, \bar{N}_{f}, 0, \frac{\tilde{N}_{c}-N_{c}}{N_{f}})\nonumber\\
\lambda_{N_{f}}&:& (N_{f}^{2}-1)\ \ (1, 1, 0,-\frac{\tilde{N}_{c}-N_{c}}{N_{f}} )
\end{eqnarray}
is dual to  
\begin{eqnarray}
\Psi_{T}&:& (1) (N_{f}, \bar{N}_{f}, 0, \frac{\tilde{N}_{c}-N_{c}}{N_{f}})\nonumber\\
B&:& (1) (1, 1, 0, \frac{\tilde{N}_{c}-N_{c}}{N_{f}})        
\end{eqnarray}
with $N_{f} = N_{c} + \tilde{N}_{c}$. We ragard that 
$\lambda_{N_{f}}$ and the ``anti-field'' of baryon $B$ combine to $(N_{f},\bar{N}_{f})(1,1,0,- \frac{\tilde{N}_{c}-N_{c}}{N_{f}})$ in the right-hand side of (3), and thus the anomaly matching in (10) and (11) is manifest. Note that the assignment of $U^{AF}_{R}(1)$ charge is arbitrary up to a common overall constant factor.

For $N_{c}=N_{f},N=1$ supersymmetric QCD, it has been argued [1] that one has a non-perturbative constraint with a QCD mass scale $\Lambda$
\begin{equation}
det T - B\bar{B} = \Lambda^{2N_{f}}
\end{equation}
where $T$ and $B$ are meson and baryon scalar multiplets, respectively, and the above duality relation corresponds to the case 
\begin{equation}
<T> = 0, \ \ <B> = - <\bar{B}> = \Lambda^{N_{f}}
\end{equation}
Namely, the baryon number is condensed in the vacuum; this explains the  peculiar assignment of vanishing baryon number to all the fields in (10) and (11). When one 
formally breaks the color symmetry as $SU(N_{f})_{c} \rightarrow SU(N_{f}-\tilde{N}_{c})_{c}\times SU(\tilde{N}_{c})_{c}$, the baryon $B$ (and also $\bar{B}$) dissociates; an assumption to this effect has been made in Ref.[2]. At the same time, one can assign  definite baryon numbers to the  
fields as in the correspondence (3) and (4). Namely, each field can pick up (basically arbitrary) baryon number from the vacuum in a way to be consistent with the anomaly matching ( or anomaly-free condition if one includes spurious ``leptons''). Note that the baryon numbers are arbitrary up to a common  overall
constant factor.  

The peculiar behavior of $U^{AF}_{R}(1)$ may also be understood in a manner similar to the spontaneously broken baryon number. The $U^{AF}_{R}(1)$ charge is 
condensed in the vacuum in the sense that the constant Grassmann parameter 
carries the charge, and each particle can pick up an arbitrary value from the 
vacuum in the above formal symmetry breaking, in a way to be consistent with anomaly condition. 

From an anomaly matching view point, we thus have two physically realizable models in one equivalence class. 
We may  picture a {\em formal} color symmetry breaking 
\begin{equation}
SU(N_{f})_{c} \rightarrow SU(N_{f}-\tilde{N}_{c})_{c}\times SU(\tilde{N}_{c})_{c} 
\end{equation}
and the duality relation
\begin{equation}
{\cal L} (Q, \bar{Q}, \lambda_{N_{f}}) \sim {\cal L} (T, B)
\end{equation}
is transformed into 
\begin{equation}
{\cal L}(\Psi_{Q}, \Psi_{\bar{Q}}, \lambda_{N_{c}}) + \bar{{\cal L}}(\Psi_{q}, 
\Psi_{\bar{q}}, \lambda_{\tilde{N}_{c}}) \sim {\cal L}(T)
\end{equation}
which in turn suggests the (electric-magnetic) duality relation
\begin{equation}
{\cal L}(\Psi_{Q}, \Psi_{\bar{Q}}, \lambda_{N_{c}}) \sim  {\cal L}(\Psi_{q}, 
\Psi_{\bar{q}}, \lambda_{\tilde{N}_{c}}) + {\cal L}(T)
\end{equation}
or 
\begin{equation}
{\cal L}(\Psi_{Q}, \Psi_{\bar{Q}}, \lambda_{N_{c}}) + \bar{{\cal L}}(T)\sim  {\cal L}(\Psi_{q}, 
\Psi_{\bar{q}}, \lambda_{\tilde{N}_{c}}) 
\end{equation}
In this picture, it is more natural to assign the color quantum number $(\bar{\tilde{N}}_{c})$ of
$SU(\tilde{N}_{c})_{c}$ to $\Psi_{q}$ as in our assignment in (2).

At this moment we have no  physical meaning assigned to the formal color symmetry breaking sequence in (14) , except for providing mnemonics for a systematic anomaly matching. Nevertheless, the peculiar behavior of baryon number and $U^{AF}_{R}(1)$ symmetry in (3) and (10) is quite suggestive and it might acquire  some significance  in the future analyses of duality. The $N_{c}= N_{f}$ case is the simplest from a view point of anomaly matching, and it is likely that  it  plays a pivotal role in the analysis.

The anomaly matching is an equivalence relation up to  a set of fields which 
are anomaly-free by themselves. It would therefore be sensible to classify the solutions of anomaly matching by  restricting
the possible set of allowed fields. We tentatively calssify a solution as a 
{\em minimal} set if we have {\em no} fields which are degenerate with respect to the quantum
numbers of $SU(N)_{c}\times SU(N_{f})_{L}
\times SU(N_{f})_{R} \times U_{B}(1)$ and  which are distinguished only by different $U^{AF}_{R}(1)$ charges. Otherwise, we 
classify  a solution as a {\em non-minimal} set.  
 As for anomaly matching,  we first impose only anomaly matching linear in $U^{AF}_{R}(1)$  and examine $U^{AF}_{R}(1)^{3}$ anomaly later.  The example  we discussed so far , i.e., $N=1$ supersymmetric QCD is classified as a minimal set. In fact we found that the minimal set in (3) automatically satisfies the $U^{AF}_{R}(1)^{3}$ anomaly matching also.     
On the other hand, the example analyzed by Kutasov and Schwimmer[8]  belongs to a non-minimal set in this classification.

\section{Non-minimal model with $SU(N)_{c}$ gauge theory}
To be specific, the model in Ref.[8] contains the following fermion contents: the starting theory is a modification of $N=1$ supersymmetric QCD with color
gauge symmetry $SU(N_{c})_{c}$ by adding  an extra chiral field $X$ in the adjoint representation of $SU(N_{c})_{c}$. In the notation of $SU(N)_{c}\times SU(N_{f})_{L}
\times SU(N_{f})_{R} \times U_{B}(1) \times U_{R}^{AF}(1)$,  we have fermion
components
\begin{eqnarray}
\Psi_{Q} &:& (N_{c})\ \ (N_{f}, 1, 1, -\frac{2}{k+1}\frac{N_{c}}{N_{f}})\nonumber\\
\Psi_{\bar{Q}}&:& (\bar{N}_{c})\ \ (1, \bar{N}_{f},- 1, -\frac{2}{k+1}\frac{N_{c}}{N_{f}})\nonumber\\
\Psi_{X}&:& (N_{c}^{2}-1)\ \ (1, 1, 0, \frac{1-k}{k+1})\nonumber\\
\lambda_{N_{c}}&:& (N_{c}^{2}-1)\ \ (1, 1, 0, 1)
\end{eqnarray}
where $k$ stands for a positive integer. 
The above multiplet is dual to the (magnetic) $N=1$ supersymmetric theory with $SU(\tilde{N}_{c})_{c}$ gauge symmetry given by (in the same notation)
\begin{eqnarray}
\Psi_{q} &:& (\bar{\tilde{N}}_{c})\ \ (\bar{N}_{f}, 1, \frac{N_{c}}{\tilde{N}_{c}}, 
-\frac{2}{k+1}\frac{\tilde{N}_{c}}{N_{f}})\nonumber\\
\Psi_{\bar{q}}&:& (\tilde{N}_{c})\ \ (1, N_{f}, -\frac{N_{c}}{\tilde{N}_{c}}, -\frac{2}{k+1}\frac{\tilde{N}_{c}}{N_{f}})\nonumber\\
\Psi_{Y}&:& (\tilde{N}_{c}^{2}-1)\ \ (1, 1, 0, \frac{1-k}{k+1})\nonumber\\
\lambda_{\tilde{N}_{c}}&:& (\tilde{N}_{c}^{2}-1)\ \ (1, 1, 0, 1)\nonumber\\
\Psi_{T_{j}}&:& (1)\ \ (N_{f}, \bar{N}_{f}, 0, 1- \frac{4}{k+1}\frac{N_{c}}{N_{f}} + \frac{2}{k+1}(j-1)), \ \ j=1....k
\end{eqnarray}
with $N_{c} + \tilde{N}_{c} = k N_{f}$. $\Psi_{T_{j}}$ stands for the fermion 
component of the $j$th  meson scalar multiplet formed of $Q(X)^{j-1}\bar{Q}$.
$\Psi_{Y}$ is a counter part of $\Psi_{X}$.   In this scheme, we have $k$ meson scalar multiplets $\Psi_{T_{j}}$ which can be distinguished only by $U^{AF}_{R}(1)$ charges, and thus classified as a non-minimal set.

This non-minimal property becomes more visible if one considers the 
 counter parts of eqs.(3) and (4) in the present example:Namely 
\begin{eqnarray}
\Psi_{Q} + \bar{\Psi}_{q}&:& (N_{c}+\tilde{N}_{c})(N_{f}, 1, 0,\frac{2}{k+1}\frac{\tilde{N}_{c}-N_{c}}{N_{f}})\nonumber\\
\Psi_{\bar{Q}} + \bar{\Psi}_{\bar{q}}&:& (\overline{N_{c}+\tilde{N}_{c}})(1, \bar{N}_{f}, 0,\frac{2}{k+1}\frac{\tilde{N}_{c}-N_{c}}{N_{f}})\nonumber\\ 
\Psi_{X} + \bar{\Psi}_{Y} + \lambda_{N_{c}} + \bar{\lambda}_{\tilde{N}_{c}}&:& (N_{c}+\tilde{N}_{c}, \overline{N_{c}+\tilde{N}_{c}})(1, 1, 0,-\frac{2}{k+1}\frac{\tilde{N}_{c}-N_{c}}{kN_{f}})\nonumber\\
\end{eqnarray}
where all the anomalies except for $U^{AF}_{R}(1)^{3}$ are matched within each
combination of fields, which is confirmed by calculations similar to (7) $\sim$ (9), and 
\begin{equation}
\Psi_{T_{j}}: (1)\ \ (N_{f}, \bar{N}_{f}, 0, 1- \frac{4}{k+1}\frac{N_{c}}{N_{f}} + \frac{2}{k+1}(j-1)), \ \ j=1....k
\end{equation}
If one recalls that $N_{c}+\tilde{N}_{c}=kN_{f}$ and the average of $U^{AF}_{R}(1)$ charge for $\Psi_{T_{j}}$ is  $\frac{2}{k+1}\frac{\tilde{N}_{c}-N_{c}}{N_{f}}$, which is relevant for anomaly matching linear in $U^{AF}_{R}(1)$, all the anomaly matching between (21) and (22) except for $U^{AF}_{R}(1)^{3}$ is almost self-evident by remembering the calculations in (7) $\sim$ (9).

It turns out that the $U^{AF}_{R}(1)^{3}$ anomaly matching is not satisfied in (21) by this minimal set of fields. There are many ways to remedy this situation. One may, for example, add
\begin{equation}
\Psi_{Z_{i}} : (1)(1, 1, 0, \alpha_{i}),  \ \ \ i=1,2,3
\end{equation}
with $\alpha_{1}+\alpha_{2}+\alpha_{3}=0$ to $\Psi_{X} + \bar{\Psi}_{Y} + \lambda_{N_{c}} + \bar{\lambda}_{\tilde{N}_{c}}$ in (21), or replace the last line 
in (21) by
\begin{equation}
(N_{c}+\tilde{N}_{c}, \overline{N_{c}+\tilde{N}_{c}}) (1, 1, 0, \beta_{i}), \ \ \ i=1,2,3
\end{equation}
with $\beta_{1}+\beta_{2}+ \beta_{3}=-\frac{2}{k+1}\frac{\tilde{N}_{c}-N_{c}}{kN_{f}}$. In either case, if one suitably chooses 3 real parameters $\alpha_{i}$
or $\beta_{i}$, one can adjust  $U^{AF}_{R}(1)^{3}$ anomaly freely by keeping 
the anomalies linear in $U^{AF}_{R}(1)$ fixed. For example, $U^{AF}_{R}(1)^{3}$ anomaly for (23) is proportional to $\alpha_{1}^{3} + \alpha_{2}^{3} + \alpha_{3}^{3}$ , which is controlled by the signature of $\alpha_{3}$ if one chooses  $\alpha_{1}\simeq \alpha_{2}$.   A characteristics of these modifications is that one needs to introduce a rather large number of color-singlet
and flavor-singlet fields, which have no clear physical meaning in the context of supersymmetric QCD. 

From an anomaly matching view point, the correspondence in (21) together with (24) suggest that $N_{f}$ flavor $SU(kN_{f})_{c}$ supersymmetric QCD, with 2 color-adjoint flavor-singlet matter 
fields added,  could be dual to $\Psi_{T_{j}}$ and 3 color-singlet flavor-singlet fields, for example. The baryon number assignment in (21) suggests the condensation of quark scalar multiplets, somewhat analogous to (13). Apparently, physical considerations other than anomaly matching are needed here to see if 
there are two physically realizable dual models, based on $SU(kN_{f})_{c}$ as well as on $SU(N_{c})_{c}$ and 
$SU(\tilde{N}_{c})_{c}$,  in this equivalence class of anomaly matching.  

\section{Discussion}
When one considers the color gauge group $SO(N)_{c}$ [2], the above 
formal color symmetry breaking in (14) is replaced by 
\begin{equation}
SO(N_{f}+4)_{c} \rightarrow SO(N_{c})_{c}\times SO(\tilde{N}_{c})_{c}
\end{equation}
where $N_{f}$ stands for the number of massless quarks and $N_{c}=N_{f}+ 4 -
\tilde{N}_{c}$. 
Note that $SO(N_{f}+4)_{c}$ theory is known to be a counter part of $SU(N_{f})_{c}$
 theory[3]. In this case, we consider the correspondence by denoting $SO(N)_{c}$ and $SU(N)_{f}\times U^{AF}_{R}(1)$ quantum numbers of fermionic components 
as 
\begin{eqnarray}
Q&:&(N_{f}+4)(N_{f}, \frac{N_{f}+2}{N_{f}+4}\frac{\tilde{N}_{c}-N_{c}}{N_{f}})
\rightarrow \left\{ \begin{array}{cc}
(N_{c})(N_{f}, \frac{2-N_{c}}{N_{f}})\\
(\tilde{N}_{c})(N_{f}, -\frac{2-\tilde{N}_{c}}{N_{f}})  
\end{array}\right.\nonumber\\
\lambda&:& (\frac{(N_{f}+4)(N_{f}+3)}{2})(1, -\frac{\tilde{N}_{c}-N_{c}}{N_{f}+4})\rightarrow\left\{ \begin{array}{cc}
(\frac{N_{c}(N_{c}-1)}{2})( 1,1)\\
(\frac{\tilde{N}_{c}(\tilde{N}_{c}-1)}{2})(1, -1)  
\end{array}\right.
\end{eqnarray} 
and its dual 
\begin{equation}
M^{ij}:(1)(\frac{N_{f}(N_{f}+1)}{2},\frac{\tilde{N}_{c}-N_{c}}{N_{f}}) 
\rightarrow (1)(\frac{N_{f}(N_{f}+1)}{2},\frac{\tilde{N}_{c}-N_{c}}{N_{f}})
\end{equation}
where $N_{f}+4 = N_{c} + \tilde{N}_{c}$,  and  $M^{ij}$ stands for a meson field formed of $Q^{i}Q^{j}$. [  $SU(N)_{f}\times U^{AF}_{R}(1)$ is basically $\gamma_{5}$ - symmetry]. Again note that $U^{AF}_{R}(1)$ assignment is arbitrary up to a common overall factor. 
We have a minimal set of fields for the anomaly matching of $SU(N)_{f}^{2}\times U^{AF}_{R}(1)$, $U^{AF}_{R}(1)$ - gravitational and $SU(N)_{f}^{3}$, without imposing $U^{AF}_{R}(1)^{3}$ matching,  when one looks at the correspondences in (26) along the arrows: The solution thus obtained automatically satisfies
$U^{AF}_{R}(1)^{3}$ anomaly matching also in both sides of (26) and (27), as is confirmed by explicit calculations[2].  We again have two physically realizable dual models,
based on $SO(N_{f}+4)_{c}$ as well as on $SO(N_{c})_{c}$ and $SO(\tilde{N}_{c})_{c}$,   in one
equivalence class of anomaly matching. 

As for the gauging of $R$-symmetry, it is known that it is consistent only within the framework of supergravity[9], though the gauging of $R$-symmetry is not 
inevitable in supergravity. It is interesting that $U^{AF}_{R}(1)^{3}$ anomaly matching, which physically suggests the gauging of $R$-symmetry, imposes a non-trivial constraint on the dynamics of supersymmetric QCD.   

In conclusion, we commented on a  specific aspect of  the non-Abelian 
electric-magnetic duality in $N=1$ supersymmetric QCD. It has been  shown that anomaly matching is not mysterious, but rather it exhibits certain regularity. 
  It is hoped that the present note stimulates further thinking about the non-Abelian electric-magnetic duality. 

I thank T. Yanagida and Y. Nomura for a helpful discussion and for calling 
Ref.[9] to my attention.

\end{document}